\documentclass[aps,prl,superscriptaddress,twocolumn]{revtex4-2}
\usepackage{graphics}
\usepackage{epsfig}
\usepackage{color}
\usepackage{stackrel}
\usepackage{amsfonts}
\usepackage{wrapfig}
\usepackage{pstricks}
\usepackage{pst-node}
\usepackage{bm}
\usepackage{dcolumn}
\usepackage{multirow}
\usepackage{epstopdf}
\usepackage{subfigure}
\usepackage{amssymb}
\usepackage{amsmath}
\usepackage{commath}
\usepackage{graphicx,bm}
\usepackage{verbatim}
\usepackage{booktabs}
\usepackage[para]{threeparttable}
\usepackage{etoolbox}
\usepackage{lipsum}
\usepackage{tcolorbox}

\newtcbox{\mywboxtext}{on line,colback=white,colframe=black,size=fbox,arc=3pt,boxrule=0.8pt}

\begin{document}
\title{Composite excitonic states in doped semiconductors}
%

\author{Dinh~Van~Tuan}
\affiliation{Department of Electrical and Computer Engineering, University of Rochester, Rochester, New York 14627, USA}
\author{Hanan~Dery}
\altaffiliation{hanan.dery@rochester.edu}
\affiliation{Department of Electrical and Computer Engineering, University of Rochester, Rochester, New York 14627, USA}
\affiliation{Department of Physics and Astronomy, University of Rochester, Rochester, New York 14627, USA}

\begin{abstract}
We present a theoretical model of composite excitonic states in doped semiconductors.  Many-body interactions between a photoexcited electron-hole pair and the electron gas are integrated into a computationally tractable few-body problem, solved by the variational method. We focus on electron-doped ML-MoSe$_2$ and ML-WSe$_2$ due to the contrasting character of their conduction bands.  In both cases, the core of the composite is a tightly-bound trion (two electrons and valence-band hole), surrounded by a region depleted of electrons. The composite in ML-WSe$_2$ further includes a satellite electron with different quantum numbers. The theory is general and can be applied to semiconductors with various energy-band properties, allowing one to calculate their excitonic states and to quantify the interaction with the Fermi sea. 
\end{abstract}

\maketitle

Optical transitions in low-temperature doped semiconductors allow us to study many-body phenomena through the interaction between photoexcited electron-hole pairs and the Fermi sea \cite{Haug_PQE, SchmittRink_AP89, Bauer_PRB92, HaugBook, Astakhov_PRB00, Combescot_Book}. For half a century, various theoretical models have been proposed to understand various observations. First studied during the late 1960s, Mahan predicted singularity in the optical conductivity due to interaction of Fermi-surface electrons with an infinite-mass valence-band (VB) hole \cite{Mahan_PR67a,Mahan_PR67b}. Back then, this problem had direct connection with  its contemporary $x$-ray edge problem due to excitation of deep core electrons in metals \cite{Anderson_PRL67,Nozieres_PR69,Schotte_PR69,JP71}.  Two decades later, the problem was extended to two- and one-dimensional electron gases in semiconductor systems \cite{Skolnick_PRL87,Hawrylak_PRB91,Kane_PRB94,Brown_PRB96,Gogolin_Book,Mkhitaryan_PRL11}, along with the idea of shakeup processes \cite{Swarts_PRL79,Sooryakumar_SSC85,Chang_PRB85,Sooryakumar_PRL87}. 

Common to these early studies was the assumption of a  relatively large Fermi energy, relevant in semiconductors when $E_F \gg \varepsilon_T$, where $\varepsilon_T$ is  gained  energy from binding an exciton with free electron to form a trion. The development of semiconductor nanostructures in the 1980s and 1990s allowed researchers to study the physics of trions in the regime $E_F \lesssim \varepsilon_T$ \cite{Kheng_PRL93,Finkelstein_PRL95,BarJoseph_SST05}. It was then suggested that the bare trion, a three-body complex, becomes correlated to the electron gas with the buildup of the Fermi sea \cite{Bronold_PRB00,Suris_PSS01,Esser_pssb01,Koudinov_PRL14}. The result is a four-body composite, termed a Suris tetron, in which the trion is bound to a Fermi hole in the conduction band (CB hole)  \cite{Suris_PSS01}. Namely,  the trion and the lack of Fermi-sea electrons in its vicinity move together. Following the discovery of monolayer transition-metal dichalcogenides (ML-TMDs) in the previous decade \cite{Splendiani_NanoLett10,Mak_PRL10,Korn_APL11,Zeng_NatNano12,Mak_NatNano12,Feng_NatComm12,Jones_NatNano13}, the interest in this topic has been revived  \cite{Dery_PRB16,Sidler_NatPhys17,VanTuan_PRX17,Chang_PRB19,Glazov_JCP20,Rana_PRB20,Liu_NatComm21,Li_NanoLett22}. Borrowing from atomic systems \cite{Frohlich_PRL11,Schmidt_PRA12}, an alternative perspective to the observed behavior in ML semiconductors has been suggested.  Rather than trions, excitons are viewed as mobile impurities in the electron gas, and the consequences of their interaction with the Fermi sea are repulsive and attractive Fermi polarons \cite{Efimkin_PRB17,Efimkin_PRB18,Chang_PRB18,Fey_PRB20}. The latter is the equivalent of a trion. 

This Letter describes a computational scheme through which many-body correlated excitonic states are converted to tractable few-body problems, solved by the variational method. The results shed light on optical measurements in ML semiconductors, providing support for the existence of excitonic states with a trion at their core surrounded by CB holes and satellite electrons if the electronic band structure supports it. Given the discrepancy with the  Fermi-polaron picture, we find it important to emphasize from the outset why it is a trion rather than exciton at the core of these states. While the exciton binding energy is an order of magnitude larger than that of the trion, $\varepsilon_X \gg \varepsilon_T$, their spatial extents do not differ appreciably because the hole is equally and strongly bound to either of the two electrons \cite{Mayers_PRB15,Kylanpaa_PRB15,Kidd_PRB16,Donck_PRB17,Mostaani_PRB17,VanTuan_PRB18}. The trion is glued by short-range forces between its three particles, where the total gained energy is $\varepsilon_X + \varepsilon_T$. The  long-range dipolar force between an exciton and electron plays a secondary role.  As such, and albeit $\varepsilon_X \gg \varepsilon_T$, it is misleading to think of a trion as a tightly-bound exciton that is loosely held together with a satellite electron. The three particles of the trion remain strongly bound as long as the average distance between electrons in the Fermi sea exceeds the radius of the trion.

\begin{figure}
\includegraphics[width=8.5cm]{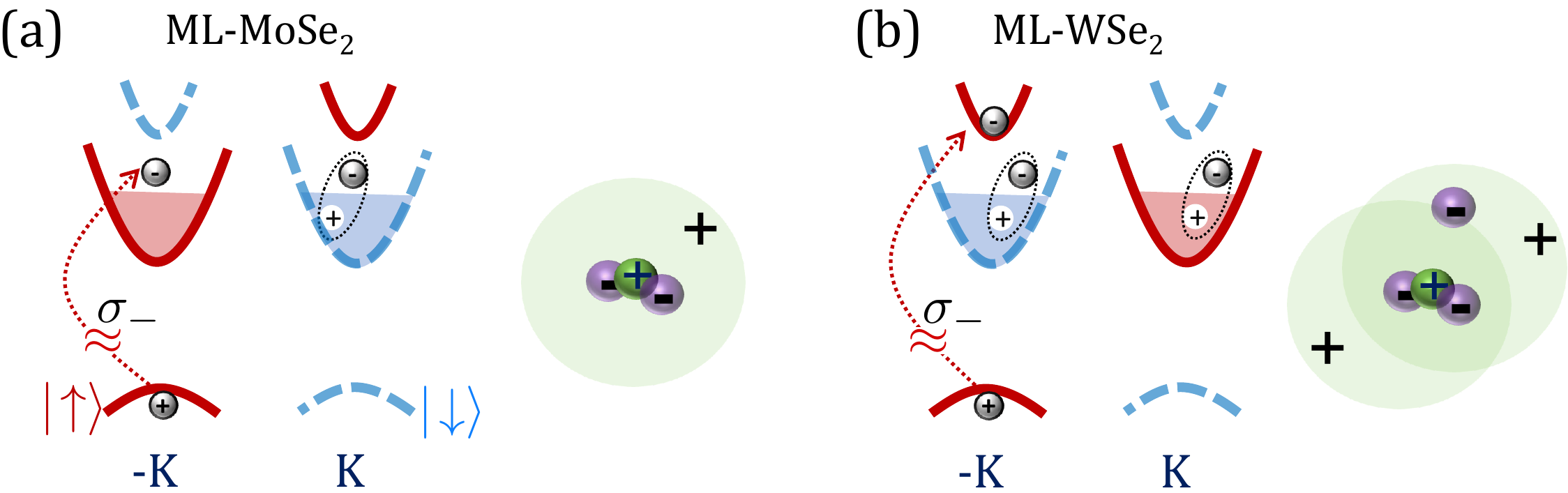}
 \caption{Composite excitonic states in electron-doped ML-MoSe$_2$ and ML-WSe$_2$ following circularly-polarized photoexcitation in the $-K$ valley. The left diagrams in (a) and (b) show the corresponding $k$-space configurations. The right diagrams correspond to real-space configurations, showing a core trion surrounded by CB hole(s). The composite in ML-WSe$_2$ is further accompanied by a satellite electron. Each electron in these composites comes with distinct spin and valley quantum numbers. 
} \label{fig:scheme}
\end{figure}

To make the discussion intelligible, we will focus on optical transitions in ML-MoSe$_2$ and ML-WSe$_2$ on accounts of their  archetypal CB structures \cite{Scharf_JPCM19,VanTuan_PRB19}. Figure~\ref{fig:scheme}(a) shows the photoexcitation process in electron-doped ML-MoSe$_2$, where the core trion is accompanied by a CB hole. The two opposite-spin electrons exhaust the  wavevector space above the Fermi surface, $k \geq k_F$, allowing them to orbit and bind the VB hole. On the other hand, the CB hole exhausts the $k$-space below the Fermi surface, $k<k_F$, and its spatial extent is commensurate with $1/k_F$. Hereafter, the combined core trion and CB hole  is referred to as a correlated trion (or tetron). Its photoexcitation is accompanied by formation of exchange-hole around the photoexcited electron, caused by exchange interaction between the photoexcited electron and electrons in $-K$ (band-gap renormalization). The electron from the time-reversed valley at $K$ is pulled out of the Fermi sea, resulting in a CB hole. This behavior is qualitatively similar to the one found in GaAs-based quantum wells in the sense that the spin and valley quantum numbers of the photoexcited electron are similar to those of electrons in the Fermi sea.   

Electron-doped ML-WSe$_2$ is different. As shown in Fig.~\ref{fig:scheme}(b), the spin-valley quantum numbers of the photoexcited electron are distinct, allowing for the generation of  a six-particle composite. The trion at its core comprises the VB hole and pulled-out electrons from the Fermi seas of the time-reversed valleys. The VB hole prefers binding tightly to these two electrons on accounts of their heavier mass compared with that of the optically-active electron in the top valley \cite{Yang_PRB22,VanTuan_PRB18}. The core trion is accompanied by two CB holes and  the `satellite' photoexcited electron. The latter captures the electron-depleted region surrounding the core trion. When the electron density increases, the radius of the depleted region shrinks ($\propto 1/k_F$), resulting in tighter binding of the top-valley electron to the rest of the complex. Hereafter, the six-particle composite is referred to as hexciton. 


Before embarking on the theory,  we mention that this Letter is part of a tetrad \cite{h,l,g}. The first study that accompanies this Letter is analysis of magneto-reflectance data in ML-WSe$_2$ \cite{h}. We show that trions evolve to hexcitons and then to 8-body composites (oxcitons) when $E_F$ crosses to the top valley of the CB.  In addition, we analyze  photoluminescence data and identify central and secondary optical transitions of hexcitons.  The evidence we provide weakens our previous argument that exciton interaction with shortwave plasmons stands behind the observed  optical transitions in electron rich ML-WSe$_2$  \cite{Dery_PRB16,VanTuan_PRX17,Scharf_JPCM19,VanTuan_PRB19}.  The second study that accompanies this Letter  focuses on computational details of the theory \cite{l}, meant to help interested readers utilize the computational model and revisit similar physics in both nascent and good-old semiconductors.  The third study is a comprehensive analysis of correlated trions and hexcitons in ML-TMDs, where we further elaborate on the screened interaction with the electron gas \cite{g}. This Letter is the center piece of the theory, which we present next.

To account for  the filling factor of Fermi-sea electrons, we use second quantization and write the Hamiltonian in momentum space ($\hbar =1$)
\begin{eqnarray}
H&=&K+V=\sum_{{\bf k}_\alpha} \frac{k^2}{2 m_\alpha} c^\dagger_{{\bf k}_\alpha} c_{{\bf k}_\alpha} \nonumber \\
&+&\frac{1}{2} \sum_{{\bf k}_\alpha, {\bf p}_\beta,{\bf q}} V_{\alpha,\beta}({\bf q}) c^\dagger_{{\bf k}_\alpha + \bf{q}} c^\dagger_{\bf{p}_\beta - \bf{q}} c_{{\bf p}_\beta} c_{{\bf k}_\alpha} \,.\label{eq:H}
\end{eqnarray}
$c^\dagger_{{\bf k}_\alpha}  $ ($c_{{\bf k}_\alpha}$) is the creation (annihilation) operator of an electron with momentum $\bf k$, and the index $\alpha$ encompasses the band index, spin, and valley quantum numbers. $V_{\alpha,\beta}({\bf q})$ is the Coulomb potential. %
Excitonic states are found from solutions of 
\begin{equation}
HC=EOC\,.
\label{Eq:secular}
\end{equation}
$H_{ij} = \langle \phi_i| K+V | \phi_j \rangle $ and $O_{ij} = \langle \phi_i|  \phi_j \rangle $ are energy and overlap matrix elements, where $\phi_i$ and $\phi_j$ are basis states and $C$ is a coefficients vector. Excluding the energy pocket of the photoexcited electron, CB electrons are assumed to be hosted in additional $N$ energy pockets across the Brillouin zone, each with distinct spin-valley configuration. The resulting basis states of the composite read 
\begin{eqnarray}
|  \phi_i\rangle  =   \sum_{\mathbf{X} } \phi_i(\mathbf{X})  \,  c^\dagger_{\mathbf{k}_0 }c_{v,\mathbf{p}_0}  c^\dagger_{\mathbf{k}_1} c_{\mathbf{p}_1} ... c^\dagger_{\mathbf{k}_{N}} c_{\mathbf{p}_{N}}  | \phi_0 \rangle \, . \,\,\, \,\,\, \,\,\,
\label{eq:basis}
\end{eqnarray}
$|\phi_0 \rangle$ is the ground state of the system (filled electronic states up to the Fermi energy, $E_F$), the photoexcited  VB hole comes from the first creation operator ($c_{v, \mathbf{p}_0}$), and $\mathbf{X} = \{ \mathbf{k}_0, \mathbf{k}_1, \mathbf{p}_1, \mathbf{k}_2, \mathbf{p}_2, ... , \mathbf{k}_{N},\mathbf{p}_{N}\}$ where the wavevector index embodies the unique valley-spin configuration of an electron in the complex. The pairs $\mathbf{k}_{i}$ and $\mathbf{p}_{i}$ denote a pulled-out electron and its CB hole. $\mathbf{k}_0$ is the wavevector of the photoexcited electron. In addition, $\mathbf{p}_0 =   \mathbf{Q} -  \mathbf{k}_0 - \sum_\ell (\mathbf{k}_\ell - \mathbf{p}_\ell)$ where $\mathbf{Q} $ is the translation wavevector of the composite (constant of motion). The overlap matrix element is then 
\begin{eqnarray}
O_{ij}  =  \langle \phi_i|  \phi_j \rangle  =  \sum_{\mathbf{X} }\,  \phi_i^{\ast}(\mathbf{X})  \, \phi_j(\mathbf{X})  \,F(\mathbf{X})  \,,
\label{eq:overlap}
\end{eqnarray}
where $ \phi_{i(j)}(\mathbf{X}) $ are basis functions and the filling factor,
\begin{eqnarray}
F(\mathbf{X})  =  f_{v, \mathbf{p}_0 }  (1-f_{\mathbf{k}_0}) \, \Pi_{\ell=1}^{N} (1-f_{\mathbf{k}_\ell})  f_{\textbf{p}_\ell}  \,,\,\,\,\,\, \label{eq:F}
\end{eqnarray}
is denoted in terms of Fermi distributions  of electrons in the filled VB ($f_{v,\mathbf{p}_0}=1$) and CB energy pockets.  

As is often the case, the computation becomes intractable already for small values of $N$. To circumvent this impasse, we first set the filling factor to $F(\mathbf{X}) =1$, and later we introduce the needed corrections to account for the $k$-space restrictions imposed by the Fermi distributions.  The motivation for this approach is that the multivariable integration  over $\mathbf{X}$ can be carried  analytically. We demonstrate the procedure by using Gaussian basis functions, $\phi_i(\mathbf{X}) =  \exp(-\tfrac{1}{2} \mathbf{X}^T \mathbf{M}_i \mathbf{X})$, where $M_i$ is symmetric, real and positive definite matrix of size $(2N+1)\times(2N+1)$. Setting $F(\mathbf{X}) =1$, the overlap matrix elements become
\begin{eqnarray}
O_{ij} =  \left( \frac{A}{4\pi} \right)^{\!2N+1} \!\frac{1}{|M|} \,\,\,, \label{eq:overlap2}
\end{eqnarray}
where $A$ is the area of the 2D system, $M=(M_i+M_j)/2$, and $|M|$ is its determinant. Focusing on the limit that $Q=0$, in which the complex resides in the light cone, the kinetic-energy matrix element due to relative motions of particles in the complex is 
\begin{eqnarray}
K_{ij} =  \left( \frac{S_w}{2m_v} + \frac{w_{0}}{2m_{e}} + \sum_{\ell=1}^{N} \frac{w_{\ell }- w_{\overline{\ell}} }{2m_\ell} \right)  \cdot O_{ij}  \, . \label{eq:Kij}
\end{eqnarray}
$w_0 = W_{0,0}$, $w_\ell = W_{\ell,\ell}$, and  $w_{\bar{\ell}} = W_{\bar{\ell},\bar{\ell}}$ are diagonal elements of $W = M^{-1}$.   The kinetic energy of the photoexcited electron is linked to $w_0$, and that of the VB hole to the sum of matrix elements in $W$ ($S_w$). Their respective masses are $m_{e}$ and $m_{v}$.  Similarly, the kinetic energy of the $\ell$-th CB electron-hole pair is linked to $w_{\ell} - w_{\bar{\ell}}$, representing  the electron energy above the Fermi level minus that of the missing electron below the Fermi level.

The potential-energy matrix elements between states $i$ and $j$ are calculated from
\begin{equation}
V_{ij} = \sum_{\lambda=0,\lambda< \eta }^{2N} V_{ij}^{\lambda, \eta} + \sum_{\lambda=0}^{2N} V_{ij}^{\lambda }\,. \label{eq:Vsum}
\end{equation}\label{eq:Vij}
The first term is the interaction between two quasiparticles $\{ \lambda, \eta \}$ and the second one is the interaction between quasiparticle $\lambda$ and the VB hole. The former reads
\begin{eqnarray}
\!\!\!\!\!\!\!\!V_{ij}^{\lambda,\eta} &=&   \Bigg(\sum_{\mathbf{q}} V_{\lambda, \eta}(\mathbf{q}) \text{exp}\{ - \gamma_{ij}^{\lambda\eta} q^2/2 \}\Bigg) \cdot O_{ij}\,. \label{eq:Vij2}  \end{eqnarray}
$\gamma_{ij}^{\lambda\eta} = D_{\lambda \lambda} +  D_{\eta \eta} - D_{\lambda \eta} - D_{\eta \lambda} $ with $D= M_i - \frac{1}{2} M_i^\text{T} \,\, M^{-1} \,\, M_i $ and $M=(M_i+M_j)/2$. The equation for $V_{ij}^{\lambda}$ is the same but with $\gamma_{ij}^{\lambda} = D_{\lambda \lambda}$.   The potential $V_{\lambda, \eta}(\mathbf{q})$ is bare or screened, depending on the identity of the involved particles $\lambda$ and $\eta$. The interaction between the three particles of the core trion are described by the bare (unscreened) Coulomb interaction because Fermi-sea electrons cannot screen their fast relative motion as long as $a_Tk_F \lesssim 1$, where $a_T$ is the trion radius. The resulting matrix element is
\begin{equation}
V_{ij}^{\lambda \eta} =   \frac{e_\lambda e_\eta}{2 \epsilon_b r_0} e^{-x} \left[ \pi \text{Erfi}( \sqrt{x})  - \text{Ei}(x) \right] O_{ij}, \label{eq:matrix_KRP}
\end{equation}
where $e_{\lambda(\eta)}$ is the charge of quasiparticle $\lambda~(\eta)$, $r_0$ is the polarizability of the 2D semiconductor, $\epsilon_b$ is the dielectric constant of the barriers around the semiconductor, and $x= \gamma^{\lambda \eta}_{ij} / 2 r_0^2$. $\text{Erfi}(x)$  and  $\text{Ei}(x)$ are imaginary error function and exponential integral functions, respectively. 

Other interactions, such as those between the core-trion particles and CB holes (or satellite electron) are weakly screened due to suppressed density fluctuations in the charge-depleted region around the core trion \cite{g}. Elaborate analysis is provided in Ref.~\cite{l}, including analytical expressions for matrix elements of the screened potential, of the exchange interaction between an electron and its CB hole, and of the band-gap renormalization (BGR) of the photoexcited electron.



Finally, the simplification made by setting $F(\mathbf{X}) =1$ is counteracted by introducing the potentials
\begin{eqnarray}
U_\ell(\mathbf{p}_\ell) &=& \left(V_e - \frac{ p_\ell^2 }{2m_{\ell}} \right)\Theta(k_F - p_\ell) \,\,, \nonumber \\ U_{\overline{\ell}}(\mathbf{k}_\ell) & = &  \left(V_{\overline{e}} +  \frac{ k_\ell^2}{2m_{\ell}} \right)  \Theta(k_\ell - k_F)\,,\,\,\,\,\,\,\, \label{eq:steps}
\end{eqnarray}
for the $\ell^{\text{th}}$ electron and CB hole, respectively, where $\Theta(q)$ is Heaviside step function.  In addition to `fixing' the kinetic energies, the energy constants, $V_e$ and $V_{\overline{e}}\,$, are chosen large enough, so that the energy minimization process avoids solutions in which the  $\ell^{\text{th}}$ electron penetrates the Fermi sea and its CB hole floats above the sea \cite{footnote_basis_number_vs_V0}.  The correction matrix elements read
\begin{eqnarray}
\!\!\!\!\!\!\!\!\!\!\!\!\!\!\!\! U_{ij}  &=&   \Bigg(   \sum_{\ell=1}^{N}  \left(1 - e^{-\beta_\ell}\right)V_e    + e^{-\gamma_\ell} \, V_{\overline{e}}  \nonumber \\  && \qquad \quad   \,\,+ \,\,
\frac{ w_{\,\overline{\ell}} g_{\,\overline{\ell}}  } {2m_\ell} \,\, - \,\, \frac{ g_{\ell}  w_\ell } {2m_\ell} \, \Bigg) \cdot   O_{ij} \,,\,\,\, \label{eq:Uij}
\end{eqnarray}
where $\beta_\ell = k_{F,\ell}^2/w_\ell$, $\gamma_\ell = k_{F,\ell}^2/w_{\overline{\ell}}$, $g_{\ell}= 1 - e^{-\beta_\ell}(1+\beta_\ell) $ and $g_{\,\overline{\ell}}= e^{-\gamma_\ell}(1+\gamma_\ell)$. $k_{F,\ell}$ is the Fermi wavenumber at the  $\ell^{\text{th}}$ energy pocket. 

The advantage of this computational method is that we work with small matrices instead of unwieldy multivariable integrals over the components of $\mathbf{X}$. One can then find the wavefunction of the composite, 
\begin{eqnarray}
\Psi(\mathbf{X}) =  \sum_i C_i \exp(-\tfrac{1}{2} \mathbf{X}^T \mathbf{M}_i \mathbf{X})\,,\,\,\,\,\label{eq:psi2N}
\end{eqnarray} 
where the coefficients $C_i$ and energy of the system are found by treating all elements of matrices $M_i$  as variational parameters \cite{l,Varga_PRC95,Mitroy_RMP13}.

\begin{figure*}
\includegraphics[width=18cm]{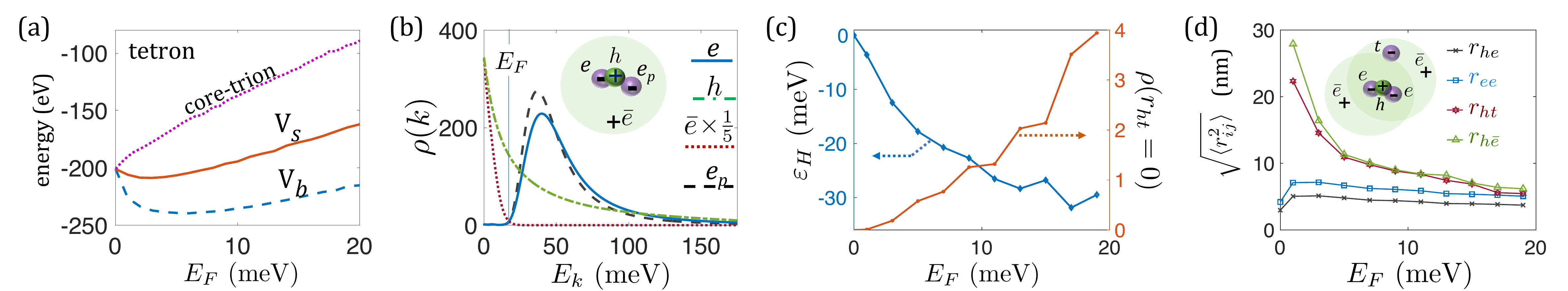}
 \caption{(a) Tetron energy vs Fermi energy ($E_F \approx 5$~meV amounts to electron density $n = 10^{12}$~cm$^{-2}$). The dotted line is the contribution from the core trion (i.e., without BGR and the CB hole). (b) $k$-space density distributions of particles in the tetron when $E_F = 19$~meV. (c) Left: Binding energy of the satellite electron to the hexciton. Right: Overlap between VB hole and satellite electron. (d) Inter-particle distances in the hexciton, showing average distances within the core trion ($r_{he}$ \& $r_{ee}$), between the VB hole and top-valley satellite electron ($r_{ht}$), and between the VB and CB holes  ($r_{h\bar{e}}$).} \label{fig:sim}
\end{figure*}



We use the model to quantify the ground-state binding energies of the correlated trion (tetron) and 6-body hexciton as a function of electron density. The dielectric parameters and effective masses are modeled by assuming that ML-WSe$_2$ is encapsulated by hexagonal boron-nitride \cite{g}. The conclusions we are about to make are qualitatively similar if we were to use effective masses and polarizability of ML-MoSe$_2$.  Figure~\ref{fig:sim}(a) shows the energy of the correlated trion, calculated by using the bare (dashed line) and screened (solid line) potentials to describe the interaction between the CB hole and core trion \cite{g}. The dotted line is the binding energy of the core trion, calculated with the restriction $k > k_F$ for its electrons, but without BGR and the CB hole.  The energy decay of the core trion  is evidently faster than the band-filling effect because it is harder for the VB hole to bind with faster electrons (large $k$).  The energy decay of the correlated trion at large densities is evidently weaker (solid and dashed lines), stemming from BGR of the photoexcited electron and offset between the reduced  $k$-space of the other electron ($k>k_F$)  with increased $k$-space of its CB hole ($k<k_F$). Experiments show that energy shifts of trion optical transitions are only moderately changed when charge is added to the semiconductor \cite{Li_NanoLett22,Liu_NatComm21,Wang_NanoLett17,Wang_PRX20,Smolenski_PRL19}. Comparing this behavior with our calculations, we see that a bare potential overestimates the binding energy of the CB hole at small charge densities, whereas the use of screened potential underestimates this binding at large charge densities. Further studies are needed to describe the correct screening effect of the electron gas \cite{g}. We continue analyzing results calculated with the screened potential, bearing in mind that the discussion is qualitatively similar when using the bare potential. Figure~\ref{fig:sim}(b) shows density distributions of particles in the correlated trion. The  distributions of the electrons are not the same because the photoexcited electron is affected by BGR (dashed black line) and the other one by electron-hole exchange with its paired CB hole (solid blue line) \cite{l}. The band filling effect can be seen from their vanishing distributions below the Fermi energy. Conversely, the wavefunction of the CB hole vanishes above the Fermi energy (dotted red line). The VB hole has no $k$-space restriction (solid green line).


Figure~\ref{fig:sim}(c) shows the binding energy of the satellite electron in the hexciton composite (i.e., the ionization energy of the hexciton), calculated from the energy difference between 6- and 5-body composites. That the binding energy of the satellite electron grows with electron density is consistent with measurements of ML-WSe$_2$ \cite{Li_NanoLett22,Liu_NatComm21,Wang_NanoLett17,Wang_PRX20}, wherein the increase in electron density results in energy redshift and amplification of the dominant optical transition \cite{h}. The redshift is analogous to increased binding energy, and the amplification to stronger overlap between the VB hole and photoexcited electron. As shown by  Fig.~\ref{fig:sim}(c), our calculations corroborate this behavior. Figure~\ref{fig:sim}(d) shows average distances between particles in the hexciton. The core trion remains intact, as can be seen from the behavior of $r_{he}$ \& $r_{ee}$. The shrinkage of the CB holes when the charge density increases can be seen from the average distance between the VB and CB holes, $r_{h\bar{e}}$. The behavior of  $r_{e\bar{e}}$ and $r_{\bar{e}\bar{e}}$ is quantitatively similar \cite{g}. The shrinkage of the CB hole further attracts the satellite electron to the core trion region, as can be seen from the behavior of $r_{ht}$. 


Before concluding this work, we mention two topics that merit further investigation. The first one deals with the strong blueshift experienced by the exciton optical transition when electrons (or holes) are added to the ML \cite{Li_NanoLett22,Liu_NatComm21,Wang_NanoLett17,Smolenski_PRL19,Wang_PRX20}. To explain this behavior, we should introduce scattered states to couple correlated trions and excitons. Since  energy levels of excitons reside in the continuum of  trion states with finite kinetic energies, the result is a Fano-like resonance blueshift of the exciton optical transition.  
The second topic that merits further investigation deals with composite excitonic states in multi-valley semiconductors such as Si or Ge. The relatively large dielectric constant in bulk semiconductors renders minuscule energy differences between the binding energies of composites with $N$ and $N+1$ particles. Incorporating such materials in low dimensional systems and encapsulating them in small-dielectric constant environments are ways to enhance the binding energy and observe composites with relatively large $N$. 

In conclusion, we have presented a theory of composite excitonic states in doped semiconductors. This important feat allows us to turn a rather difficult many-body problem into a computationally manageable few-body problem, which embodies the interaction between the electron gas and excitonic complexes. Using this method, we have calculated the tetron and hexciton states in monolayer transition-metal dichalcogenides. Hopefully, the theory will help to sort out the on-going debate on the origin of optical transitions in  doped semiconductors, and will spark a search for composite excitonic states  in various multi-valley semiconductors.

\acknowledgments{This work was supported by the Department of Energy, Basic Energy Sciences, Division of Materials Sciences and Engineering under Award DE-SC0014349 (DVT), and by the Office of Naval Research under Award N000142112448 (HD). }


\begin{thebibliography}{99}
\bibitem{footnote_basis_number_vs_V0} The chosen values of $V_e$ and $V_{\overline{e}}$ should be (i)  much larger than the binding energy, and (ii) such that the basis functions can describe the step function. A larger set of basis functions allows one to choose larger $V_e$ and $V_{\overline{e}}$.
\end{thebibliography}
\end{document}